\documentclass{IEEEtran}
\usepackage{cite}
\usepackage{amsmath,amssymb,amsfonts}
\usepackage{graphicx}
\usepackage{textcomp,nicefrac}
\usepackage{graphicx}
\usepackage{subcaption}
\usepackage{lipsum}
\usepackage{lineno}

\def\BibTeX{{\rm B\kern-.05em{\sc i\kern-.025em b}\kern-.08em
T\kern-.1667em\lower.7ex\hbox{E}\kern-.125emX}}
\begin{document}
\title{Automatic test system of the back-end card for the JUNO experiment}
\author{{Barbara Clerbaux$^{1}$, Shuang Hang$^{1, 2}$, \underline{Pierre-Alexandre Petitjean}$^{1}$, Peng Wang$^{1, 3}$, Yifan Yang$^{1}$}
\thanks{Submitted for review on October 31, 2020 \\
$^{1}$Inter-University Institute For High Energies, Universi\'e libre de Bruxelles (ULB)\\
$^2$Department of Nuclear Science and Technology, Nanjing University of Aeronautics and Astronautics \\
$^3$ College of Environmental and Biological Engineering, Nanjing University of Science and Technology}}

\maketitle

\begin{abstract}
The Jiangmen Underground Neutrino Observatory (JUNO) is a medium-baseline neutrino experiment under construction in China, with the goal to determine the neutrino mass hierarchy. The JUNO electronics readout system consists of an underwater front-end electronics system and an outside-water back-end electronics system. These two parts are connected by 100-meter Ethernet cables and power cables.
The back-end card (BEC) is the part of the JUNO electronics readout system used to link the underwater boxes to the trigger system is connected to transmit the system clock and triggered signals. Each BEC is connected to 48 underwater boxes, and in total around 150 BECs are needed. It is essential to verify the physical layer links before applying real connection with the underwater system. Therefore, our goal is to build an automatic test system to check the physical link performance.
The test system is based on a custom designed FPGA board, in order to make the design general, only JTAG is used as the interface to the PC. The system can generate and check different data pattern at different speeds for 96 channels simultaneously. The test results of 1024 continuously clock cycles are automatically uploaded to PC periodically. We describe the setup of the automatic test system of the BEC and present the latest test results.
\end{abstract}

\begin{IEEEkeywords}
JUNO, BEC, Back-End Card, GCU, Global Control Unit, TTIM, FPGA, Automatic test
\end{IEEEkeywords}

\section{Introduction}
\label{sec:introduction}

The Jiangmen Underground Neutrino observatory (JUNO)  \cite{JUNO_phys} experiment is a large liquid scintillator detector aiming at measuring electron antineutrinos issued from nuclear reactors located at a distance of about 53 km. The precise measurement of the electron antineutrino energy spectrum will allow determining the neutrino mass hierarchy (NMH), and reducing the uncertainty below 1\% on solar oscillation parameters. The required energy resolution to discriminate between the normal and inverted NMH at a 3-4 $\sigma$ level for about 6 years of data taking is 3\% at an energy of 1 MeV.  

The JUNO detector is also capable of observing neutrinos/antineutrinos from terrestrial and extra-terrestrial sources, including geoneutrinos, atmospheric neutrinos, solar neutrinos, the supernova neutrinos, and diffuse supernova neutrino background. In addition, sterile neutrinos with 10$^{-5} < \Delta m ^{2}_{41} < 10^{-2}$ eV$^2$ and a sufficiently large mixing angle $\theta_{41}$  could be identified through a precise measurement of the antineutrino energy spectrum. As a result of JUNO's large size, excellent energy resolution and vertex reconstruction capability, interesting new data on these topics will be collected. More details on the science potential of JUNO can be found in Ref. \cite{JUNO_phys}. 

The option of adding a smaller detector (2.6 tons of Gd-loaded liquid scintillator, read with SiPMs), called TAO (Taishan Antineutrino Observatory) located at 40 m distance from the Taishan reactor is being considered and studied. The aim of TAO is to measure the antineutrino spectrum at  the percent level, in order to provide a model-independent reference spectrum for JUNO and a benchmark for investigating the nuclear database.

The JUNO collaboration includes about 600 scientists, from 77 institutions, in 17 countries. The international JUNO collaboration was established in 2014. The civil construction and the PMT production started in 2015 and 2016, respectively, and are expected to end in 2020. In 2021, the detector assembly and installation will take place, followed by the liquid scintillator filling. The start of the data taking is expected in 2022.

\section{The JUNO experiment}

The requirements on the energy resolution (3\% at 1 MeV) and the energy scale (better than 1\%) put strong constraints on the detector size and on its component quality. A large volume of liquid scintillator is needed, with a high attenuation length. A high photocathode coverage implies an important number of the large the photomultiplier tubes (PMTs) to be deployed. To remove background as much as possible, vertex reconstruction capabilities need to be on place. 

The JUNO detector will be located at 700 m underground, and it will consist of a central detector, an active veto system and a calibration system. The central detector is composed by a 35 m diameter transparent acrylic sphere, filled by 20 ktons of liquid scintillator (LS), being the antineutrino target. The LS consists of Linear Alkyl Benzene (LAB) as solvent, 3 g/L 2,5-diphenyloxazole (PPO) as the fluor and 15 mg/L p-bis-(o-methylstyryl)-benzene (bis-MSB) as the wavelength shifter. The stainless steel support structure holds the inner vessel and the PMTs. Dynode and micro-channel plate PMTs are used to detect the scintillation photons from inverse beta decay (IBD) events in the LS. The acrylic sphere is instrumented with more than 18000 20-inch PMT, and about 25000 3-inch small PMTs located in between the large PMTs. The 20-inch PMT's detection efficiency (the product of the quantum efficiency times the collection efficiency) is required to be 27\% on average and 24\% on minimum. The 3-inch PMTs serve as an additional standalone calorimetry which will not face any saturation effect and that will provide handle to control the non-linearity effects.  

\begin{figure}[ht!]
\centering
\includegraphics[width=3.45in]{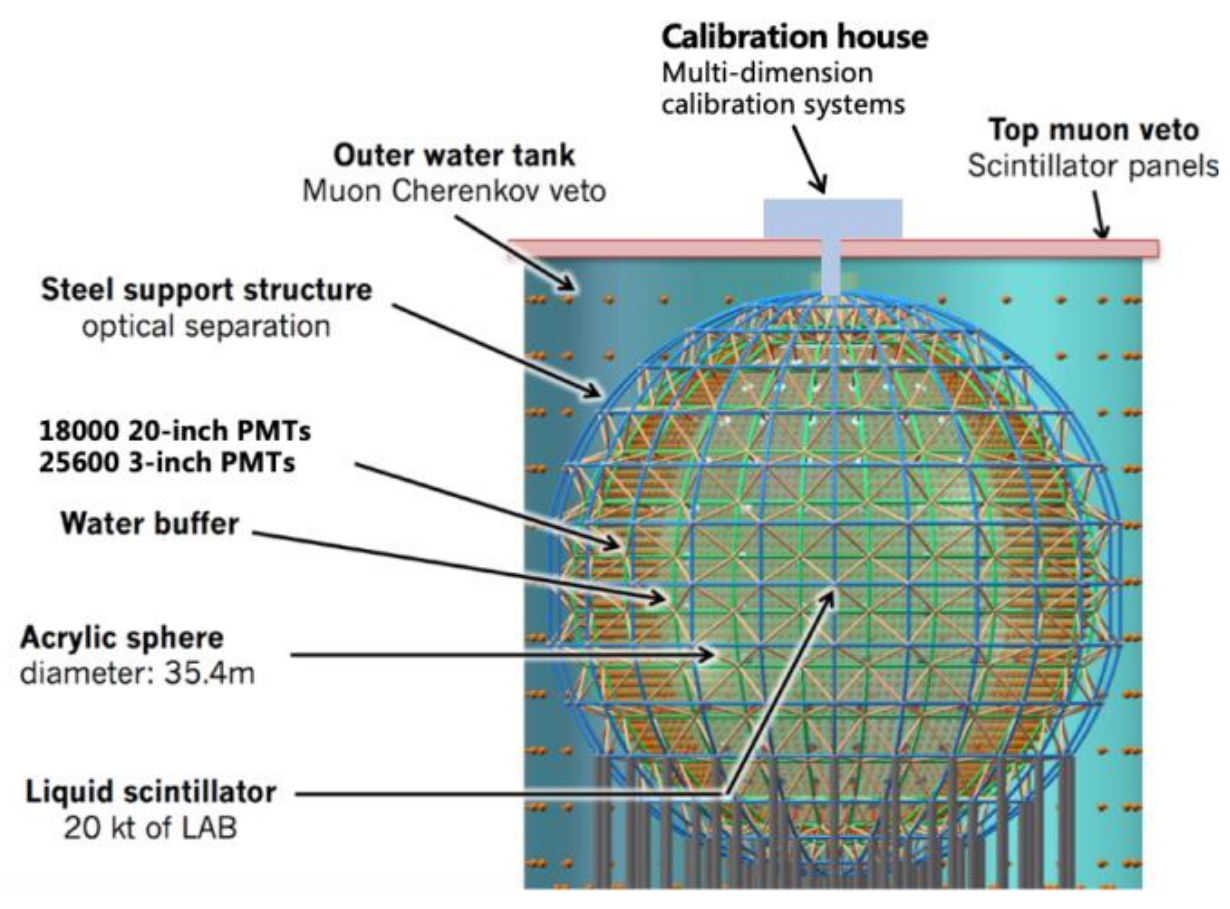}
\caption{The JUNO detector with the central acrylic sphere, the two veto systems and the calibration part.}
\label{figure:detector}
\end{figure}

Two vetoes are foreseen to reduce the different backgrounds. A 20 ktons ultrapure water Cerenkov pool around the central detector instrumented with about 2000 20-inch PMTs will tag events coming from outside the neutrino target. It will also act as a passive shielding for neutrons and gammas. In addition, a muon tracker will be installed on top of the detector (top muon veto) in order to tag cosmic muons and validate the muon track reconstruction. The top muon veto will use the target tracker detector previously used in the OPERA experiment at Gran Sasso. The JUNO detector is detailed in Ref. \cite{JUNO_det}. A schematic view of the detector is presented in Fig. \ref{figure:detector}. The fon-tend and readout electronic for the large PMT system are an important component of the experiment and their performance is crucial for the success of the various JUNO measurements. 

A sophisticated calibration complex is designed for multiple source deployments. It will also allow to cover the entire energy range of reactor antineutrinos, and the coverage of the full volume of the central detector.  Four different calibration systems will be used:  an automatic calibration unit (vertical scan), a cable loop system (complete vertical plane scan), a guide tube control system (vertical plane scan from outside of the acrylic sphere) and finally a remotely operated vehicle which will transport a source to any place in the liquid scintillator.

\section{The JUNO electronics system}

The physics requirement on the excellent energy resolution, especially at low energy, for the NMH measurement implies various specifications on the electronics: an excellent photon's arrival time measurement for a good vertex reconstruction, a large dynamic range (for atm-, geo-, and supernova neutrinos), a negligible deadtime (for supernova events lasting up to a few seconds). The following specifications have been defined: to provide a full waveform digitization with a high speed (1 Gsample/s), a high resolution (nominal resolution of 14 bits) on the full dynamic range: from 1 to 1000 pe. The main risk concerns the reliability of the underwater electronics that will be not accessible after the installation. The reliability requirement to have less than 1\% PMT and underwater electronics failure over 6 years is studied using complementary methods: a direct calculation from the various components and laboratory reliability and ageing tests.

\begin{figure}[ht!]
\centering
\includegraphics[width=3.5in]{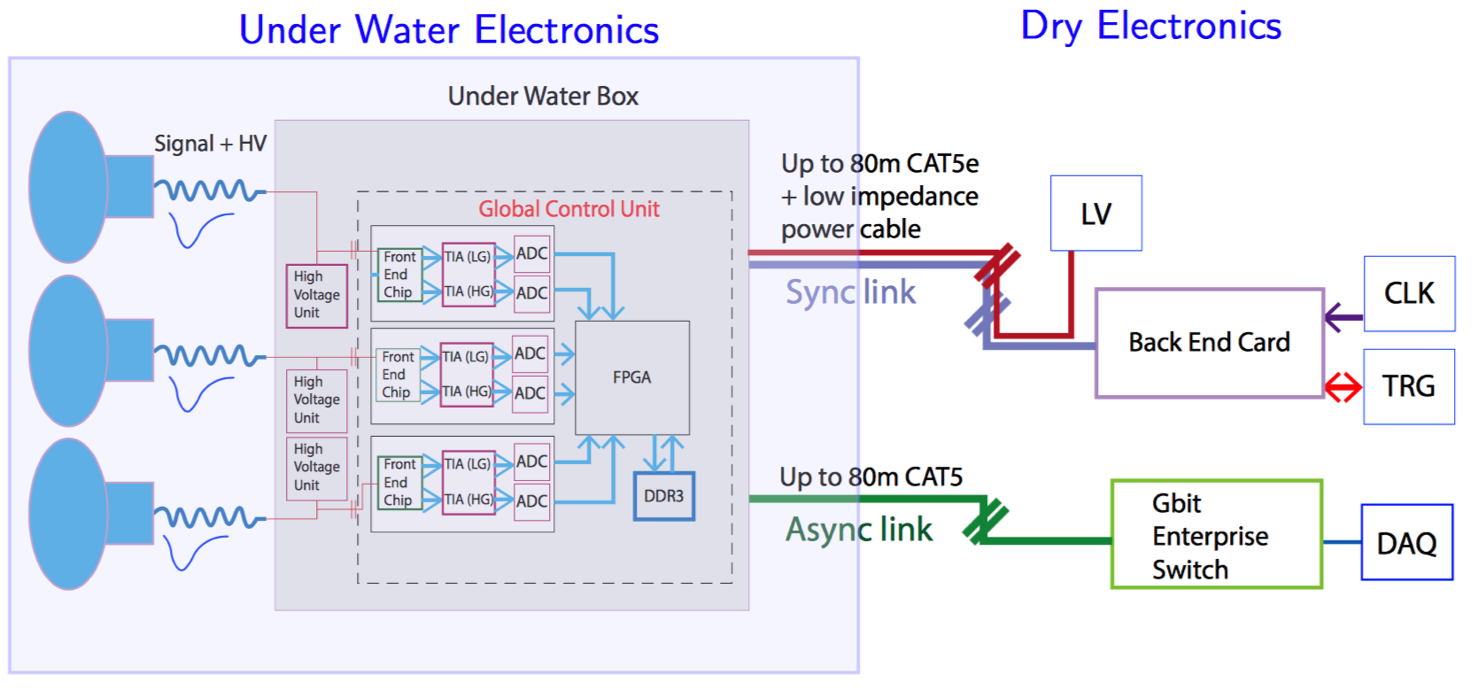}
\caption{Schematic view of the electronics readout system of JUNO : the underwater part on the left side and the above water part on the right side.}
\label{figure:electronics}
\end{figure}

The JUNO electronics system \cite{Bellato2021} can be separated into two main parts:
\begin{enumerate}
    \item the front-end electronics system, performing analog signal processing (the underwater electronics). 
\item the back-end electronics system, sitting outside water, consisting of the DAQ and the trigger. The back-end electronics system is connected to the front-end electronics system  through  100 m cables.
\end{enumerate}
A schematic view of the electronics readout system of JUNO is given in Fig. \ref{figure:electronics}.

The PMT current signal is conditioned, duplicated into two streams and converted to a voltage value (low-gain and high-gain TIAs) as follows: for the low gain (8:1) in the case of 0-1000 pe, and for the high gain (1:1) in the case of 1-128 pe. Each stream is digitized with a 14-bit 1 Gsample/s custom designed ASIC, developed by Tsinghua University. The digital signal is then processed in the FPGA (Xilinx kintex7). The reconstructed (timestamp, charge) and the digitized waveforms are then stored locally (2 GB DDR RAM). All the Global Control Units (GCUs), about 7000, are synchronised in a time window lower than $16$ ns. See Refs. \cite{GCU,Timing_Pedretti} for more details on the GCUs and the clock system, and on the timing synchronisation, respectively.

In the "Global-Trigger" running mode, a local trigger signal is sent to the Global Trigger and, if validated, data are transferred to DAQ through Ethernet. In the "Auto-Trigger" mode, fixed window waveforms (300 ns) are sent to the DAQ every time a local trigger is issued. 

The digital signal and the trigger information are forwarded to the dry electronics by means of 100 m CAT5 Ethernet cables. From each under water box, 3 cables are connected to the dry electronics: (1) one CAT5 UTP cable, which makes the variable latency ethernet link (asynchronous link) for the data readout and slow control (using the IPBUS protocol and a nominal link speed of 1Gbps);  (2) one CAT5 FTP cable, making the fixed fixed latency link (synchronous link) to the BEC for the trigger and the clock (using the Timing Trigger and Control (TTC) protocol and a nominal link speed of 125 Mbps); (3) a low impedance power cable.   

\begin{figure}[t!]
\centering
\includegraphics[width=3.5in]{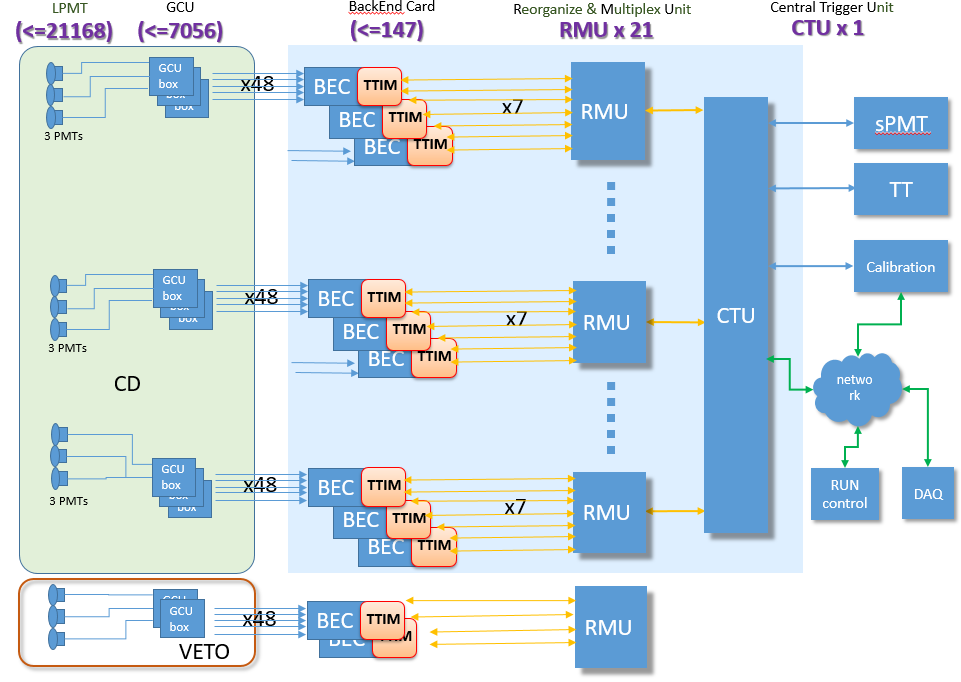}
\caption{Schematic view of the trigger system of JUNO. It is composed of the RMU (reorganise \& multiplex unit), the vertex fit logic (VFL) and the central trigger unite (CTU).}
\label{fig:Trigger}
\end{figure}

Figure \ref{fig:Trigger} shows the scheme used for the JUNO trigger. Three PMT are connected to a Global Control Unit (GCU) through a waterproof coaxial cable. 
For the back-end electronics part, the BEC are used as concentrators to collect and compensate the incoming trigger request signals. An FPGA mezzanine card handles all the trigger request signals \cite{yang2018design}. Each BEC receives 48 Ethernet cables from the 48 underwater boxes, and distributes the clock signal to the GCU's. The signals from the various BECs are sent to 21 RMU (Reorganise\&Multiplex Unit) cards, and their sum is forwarded to the CTU (Central Trigger Unit).  

The readout system of the about 25600 small (3-inch) PMTs system is similar to the one of the large (20-inch) PMTs, in particular for the outside water part. Signals from the 128 small PMTs are collected into an underwater box, which includes all the frontend electronics. On the one hand, it hosts the high-voltage power supplied and the splitting signals. On the other hand, it contains the signal readout and digitization systems, which are performed by a electronics board holding 8 CATIROC ASICs, controlled by an FPGA. The CATIROC ASIC allows working on the single photo-electron (p.e.) mode. It provides a charge measurement over a dynamic range from 1 to several hundreds of photoelectrons. The 200 underwater boxes from the small PMTs are connected to the outside water system in the same way as for the large PMT, with a total of 5 BECs needed. 


\section{The Back-End Card system}

As is shown in Figure \ref{figure:electronics}, the BEC can be seen as concentrator board. The links for the data exchange between the underwater electronics and the back-end electronics are performed through Ethernet cables (represented as blue lines in Figure \ref{figure:electronics}). Ethernet cable was chosen due to it’s high reliability and low cost.

All pairs of Ethernet cables are used. Table \ref{pairs2} gives a summary of the BEC-GCU link. Two pairs out of 4 inside the Ethernet cable are used to transfer data from the BEC to the GCU (trigger and clock running at 125 Mbps and 62.5 MHz). The other 2 pairs are used to send data from the GCU to the BEC. \\
\begin{table}[h]
\begin{center}
\begin{tabular}{|l |l |l c l| }
\hline
Name & Type of signals & origin & to & destination\\
\hline
    Pair 1-2: & 62.5 MHz clock & BEC & $\rightarrow$ & GCU\\
    Pair 4-5: &125 Mbps data & BEC &$\rightarrow$ & GCU \\
    Pair 3-6: &125 Mbps data & GCU &$\rightarrow$ & BEC \\
    Pair 7-8: &125 Mbps data & GCU& $\rightarrow$ & BEC \\
    \hline
\end{tabular}
\caption{ Connection table of the BEC-GCU link}
\label{pairs2}
\end{center}
\end{table}

\begin{figure}[ht!]
\centering
\includegraphics[width=3.5in]{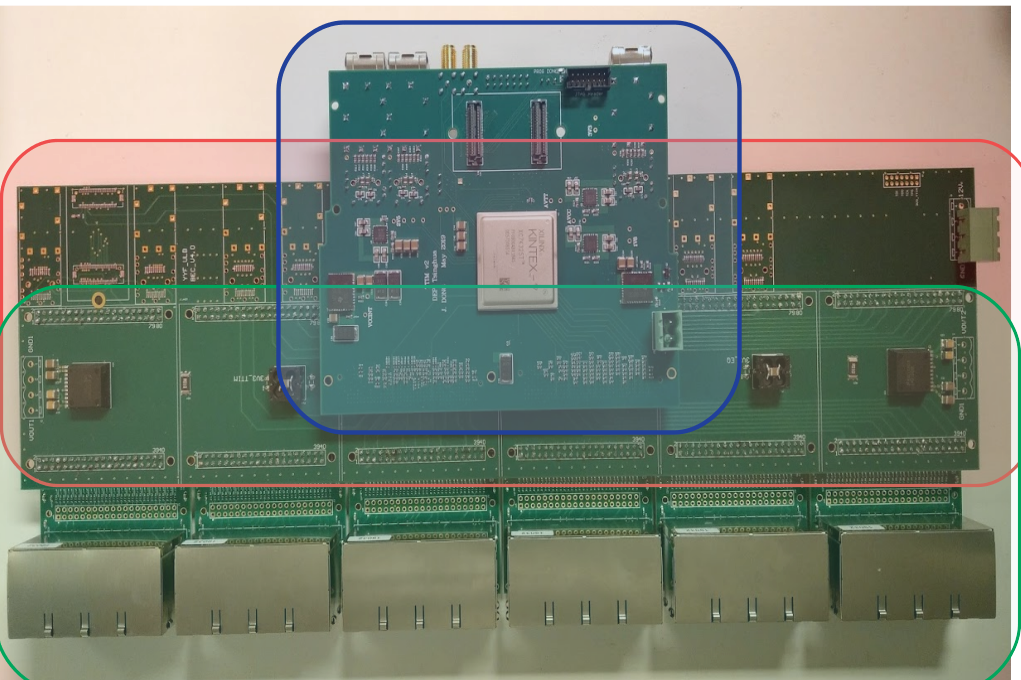}
\caption{Picture of the back-end card version 4.0. Trigger and Timing (TTIM) FMC mezzanine card is the blue box. The BEC v4 is composed of 6 mezzanine cards (green box) and one baseboard (red box). The baseboard provides power the mezzanine cards and the TTIM.}
\label{fig:BEC}
\end{figure}

Figure \ref{fig:BEC} shows the BEC and the Trigger and Timing (TTIM) FMC mezzanine card (blue box). The TTIM connects the BEC to the trigger system. The BEC v4 is composed of 6 mezzanine cards (green boxes) and one baseboard (red box). This structure was chosen because of the following reasons:

\begin{itemize}
    \item  We chose the small PCB boards as they are more reliable than larger ones and it is easier to achieve a better power distribution performance.
    \item The mezzanines are plugged to the baseboard. This makes easier to replace a defect mezzanine than changing the whole BEC. It is also more  flexible for future updates.
    \item It is more flexible to use in the case of small scale combined tests.
\end{itemize}

\section{automatic test system}

For the BEC, before participating to any combined test, a standalone test is essential. We use the prbs IP core to generate the test data stream. A VIO core is used to configure the prbs core, including prbs pattern and speed, as well as the injection of an artificial error into the data stream. The prbs errors are saved into a register. A clock counter running at the system clock is used to record the timing stamp of the first error. Both the error and the timing stamp register are accessible from a ILA core. Thus by monitoring the ILA core, we can easily get the performance of the link running during a certain time period without the need of any external connection.
We then developed a TCL based code to automatically save the whole ILA status into CSV files
periodically. By analyzing the CSV files, we are able to get the cross-section view of the link with a
resolution of about $1 Hz$, which is enough to show the behavior and trend of the link.
\\

\subsection{Test method}

\begin{figure}[ht!]
\centering
\includegraphics[width=3.3in]{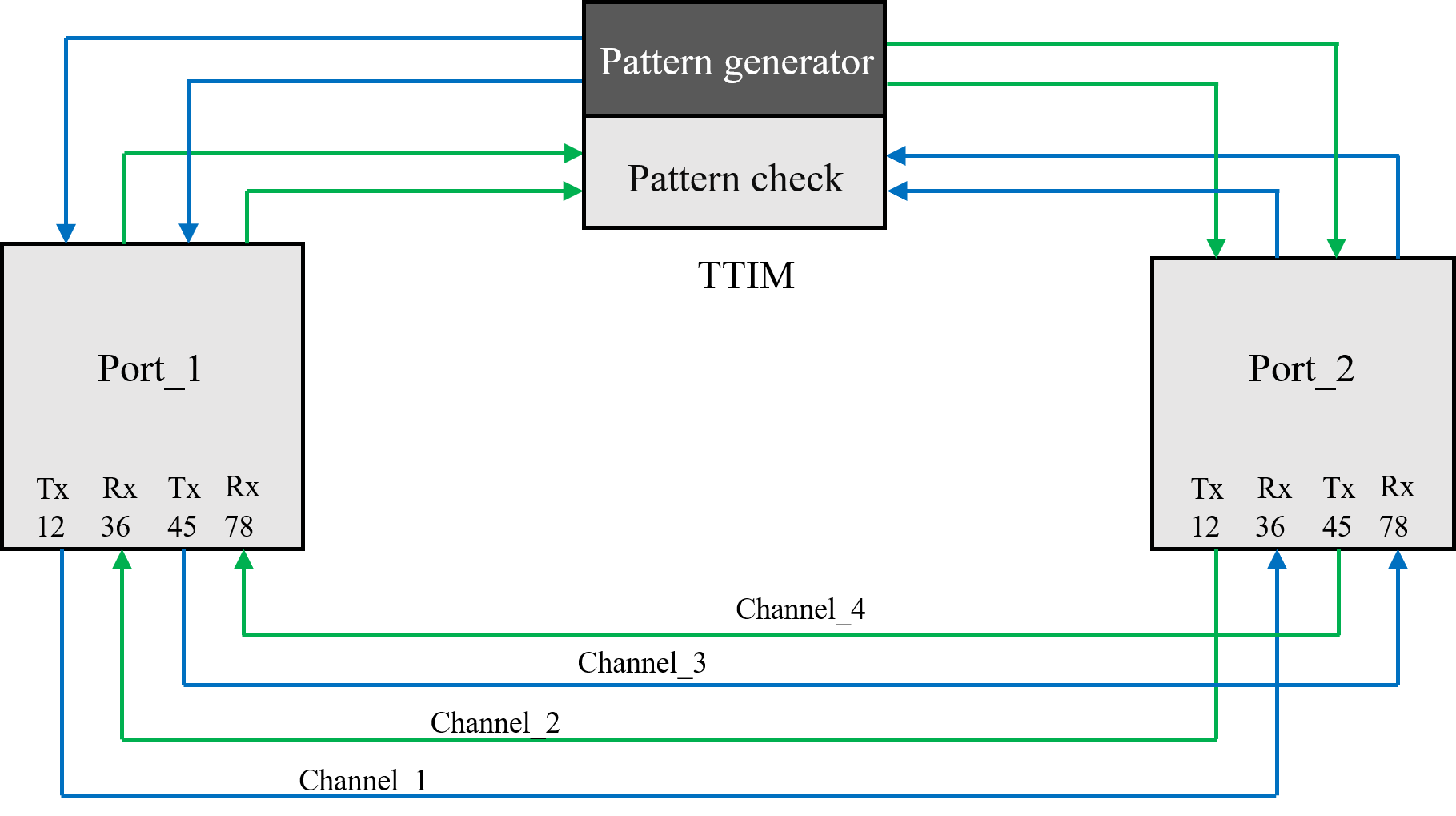}
\caption{A prbs IP core is implemented in the TTIM FPGA to generate and verify the prbs data stream. Each port is connected to a other port of the same BEC trough a Ethernet cable. In the cable we have 4 pairs, each pair is represented by a line. }
\label{fig:PRBS}
\end{figure}

Since our bandwidth is less than 10 Gbps, the prbs-7 is the commonly used data stream to verify the physical link performance. We thus take the benefit of the TTIM, by coding a prbs generator and a prbs checker on the TTIM. We are then able to establish a test link by connecting one port of the BEC to another port through 100 meters Ethernet cable. By checking the prbs error in a certain time duration, we can get the bit error rate for different links and at different speeds, as shown on Figure \ref{fig:PRBS}.

\begin{figure}[ht!]
\centering
\includegraphics[width=3.3in,height=2.0in]{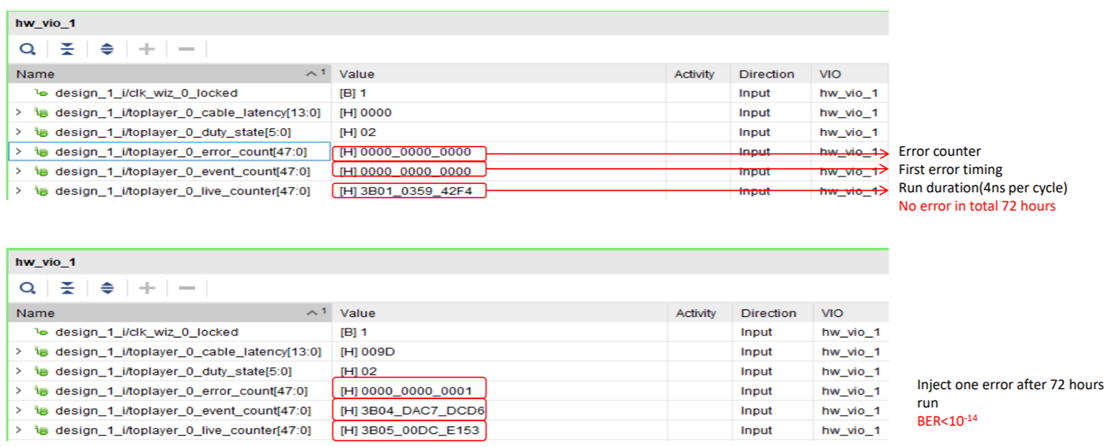}
\caption{The top image shows the VIO, used to control the error number as well as the timing stamp when a first error occurs. On the second image we inject a error after 72
hours and the VIO  correctly capture the error.  }
\label{fig:VIO}
\end{figure}

A prbs IP core is used to generate and verify the prbs data stream. A VIO and a ILA core are used to control the prbs IP core speed, and to record the error number as well as the timing stamp when the first error occurs.
Figure \ref{fig:VIO} shows the user interface of VIO. The figure shows that we are able to run the link at a speed of 250 Mbps error-freely for more than 72 hours. In order to verify the correctness of our firmware, an artificial error was injected into the data stream after 72 hours, and the VIO captured the error correctly while recording the corresponding timing stamp. 

\subsection{Setup for long-term test}
 
\begin{figure}[ht!]
\centering 
  \includegraphics[width=3.0in]{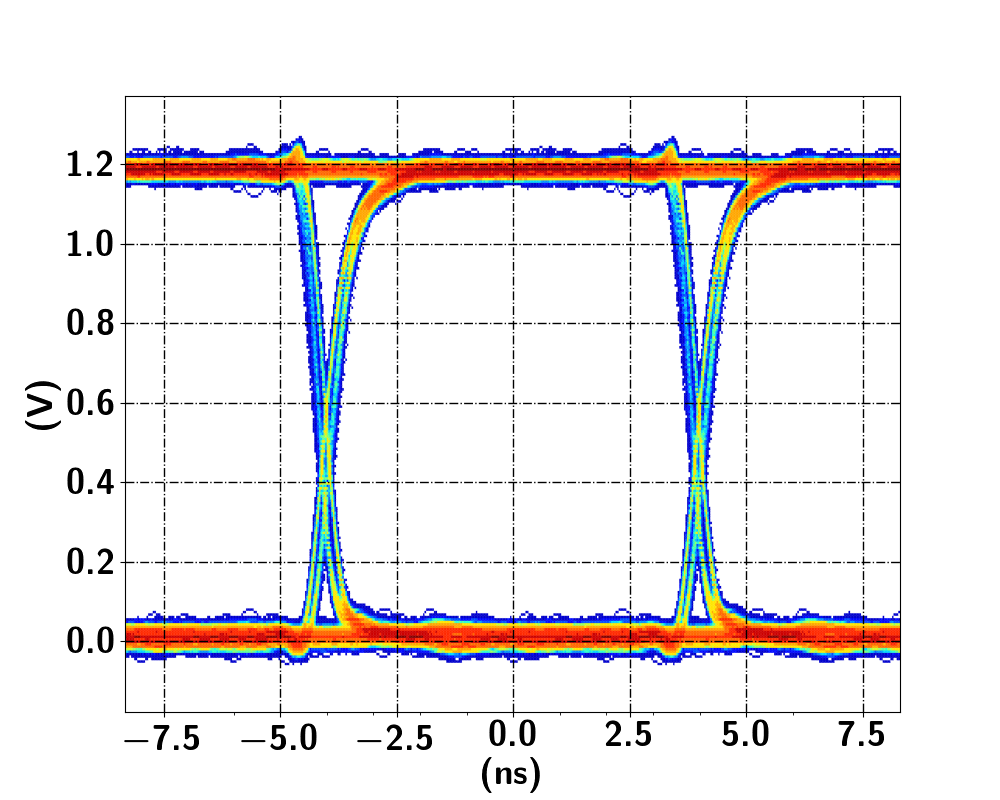}
    \caption{Eye diagram for 100 m CAT5E Ethernet cable. This eye diagram is performed for each channel of the BEC before running the TCL code. This helps to ensure that the channel is not noisy or have any connection error. The width is equal to 6.531 ns for the 100 m CAT5E Ethernet cable. We can easily achieve error-free run by adjusting the sample point.}
  \label{fig:eye}
  \end{figure}

 Figure \ref{fig:eye} shows the eye diagram for a 100 m CAT5E cable.
 Before running the long-term test, we checked the eye diagram of each channel to make sure that there was no glitch. The typical width of the eye is around 6.5 ns for CAT5E. We can easily achieve an error-free run by adjusting the sample point.
 
\subsection{Results of the BEC long-term tests}

\begin{figure}[ht!]
\centering
\includegraphics[width=2.6in]{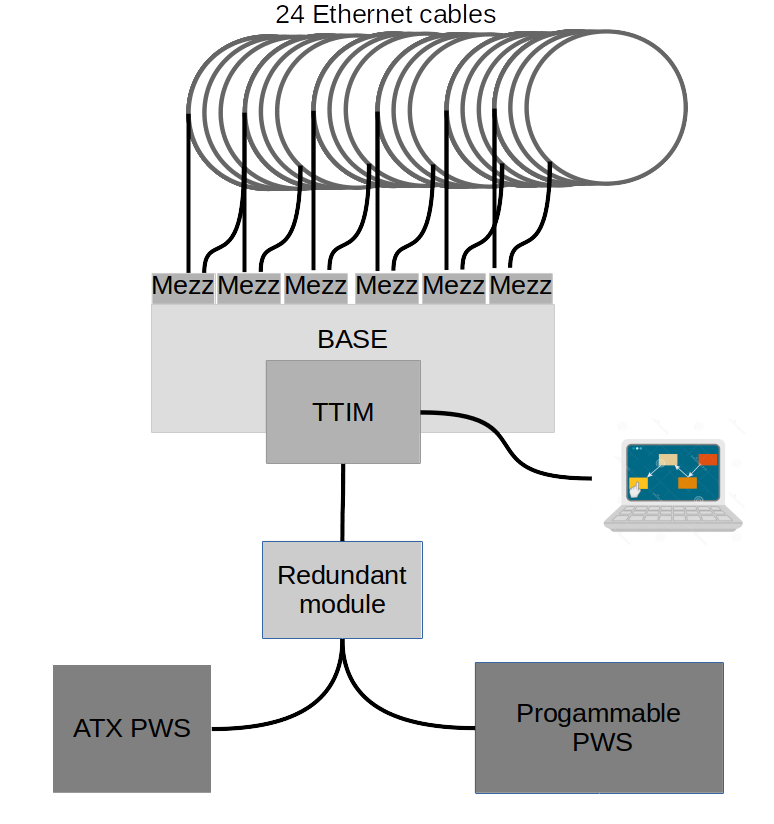}
\caption{Schematic view of the setup used to performed the BEC long-term test. The BEC is powered by a redundant power module connected to the ATX power supply (ATX-PWS) and a programmable power supply. The setup is built to change the power source every second (i.e. the BEC is 1 second powered by ATX power supply and 1 second by the programmable power supply). }
\label{fig:setup}
\end{figure}

The setup used to performed the BEC long-term test is shown in Figure \ref{fig:setup}. The BEC is powered by a redundant power module connected to the ATX power supply (ATX-PWS) and a programmable power supply. The BEC test has been running for 28 days.

\begin{figure}[t!]
   \begin{center}
  \begin{subfigure}{0.75\columnwidth}
  \includegraphics[width=\textwidth]{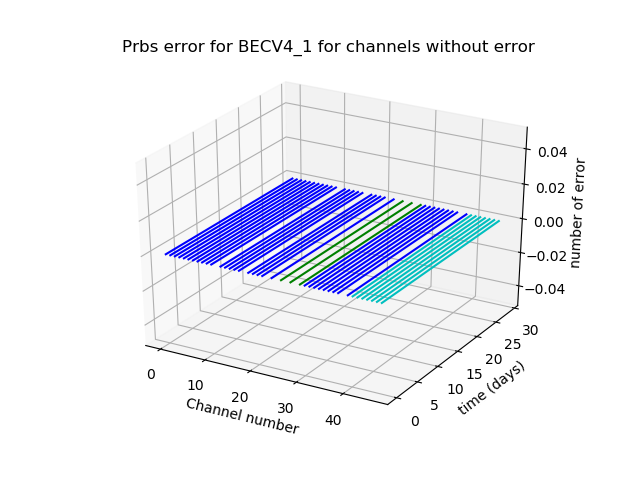}
  \caption{}
  \end{subfigure}
  \hfill
  \begin{subfigure}{0.55\columnwidth}
  \includegraphics[width=\textwidth]{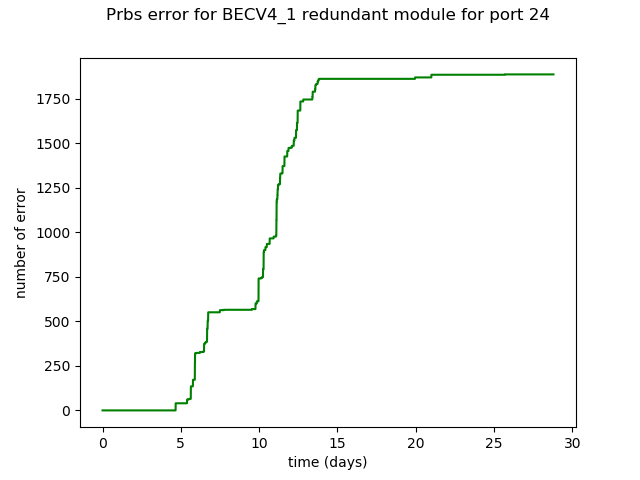}
  \caption{} 
  \end{subfigure} 
  \begin{subfigure}{0.55\columnwidth} 
  \includegraphics[width=\textwidth]{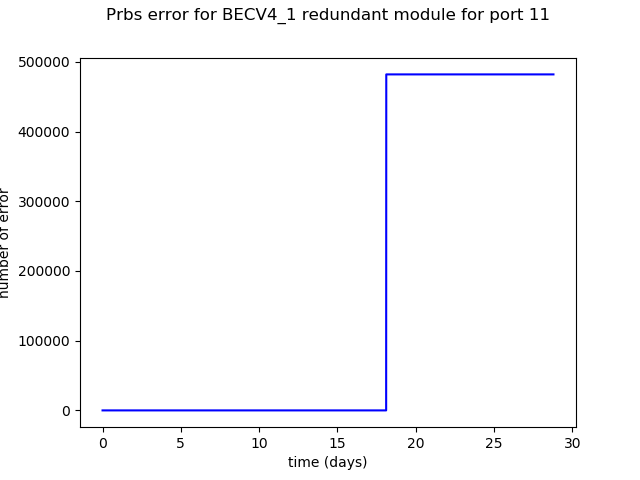} 
  \caption{} 
  \end{subfigure}  
  \hfill 
  \begin{subfigure}{0.55\columnwidth} 
  \includegraphics[width=\textwidth]{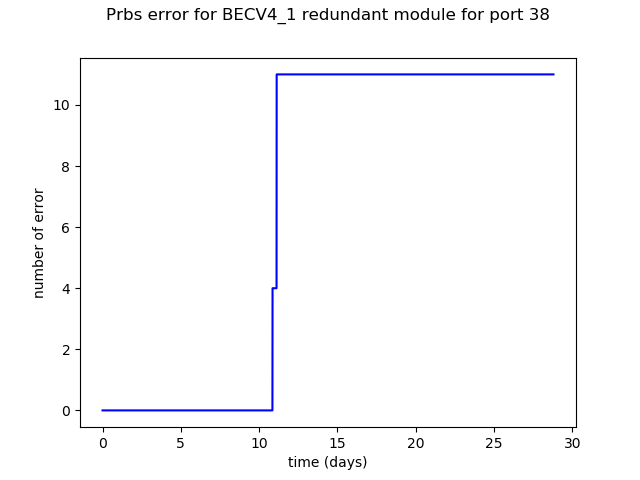} 
  \caption{} 
  \end{subfigure}
  \caption{Test results for 28 days. The upper panel (a) shows the results for the 41 channels that have no error for 28 days of running. The different colors correspond to different cables: dark blue is for 10 m CAT5E cables, light blue for 100 m CAT5E cables and green is for CAT6 cables. We have 7 channels with errors (channel numbers 11, 17, 22, 24, 26, 28 and 38). On the second panel (b) we represented the channel number 26.  The two lower panels (c) and (d) show some examples of the 6 other channels where errors occur  at a specific time, coming from an external noise.}
  \label{fig:result}
    \end{center}
\end{figure}

Figure \ref{fig:result} shows the results of the test for the 48 channels, for the 28 days of running. The upper panel (a) shows the results for the 41 channels that have no error during the 28 days of running. The different colors correspond to different cables: dark blue is for 10 m CAT5E cables, light blue for 100 m CAT5E cables and green is for CAT6 cables. We have 7 channels with errors (channel numbers 11, 17, 22, 24, 26, 28 and 38). On the second panel (b), we represent the channel number 26. Errors are appearing at different specific times, so the number of errors is increasing with the time. To debug this error, the cable was changed and no more error was seen. The two lower panels, (c) and (d), show some examples of the 6 other channels where errors occur. For those channels, the error occurred only at a specific time. This type of error comes from an external noise.

\subsection{Error capture}
During the long term test, sometimes there are errors, and it is important to find out the source of the errors before finding a possible solution. Thus, it is essential to capture the error and the behavior of the related hardware. For this, we need first to determine if the error comes from the data source or from the external noise. Therefore we setup a platform combining both hardware and firmware to track down the behavior all over the data chain.

\begin{figure}[t!]
\centering
\includegraphics[width=2.15in]{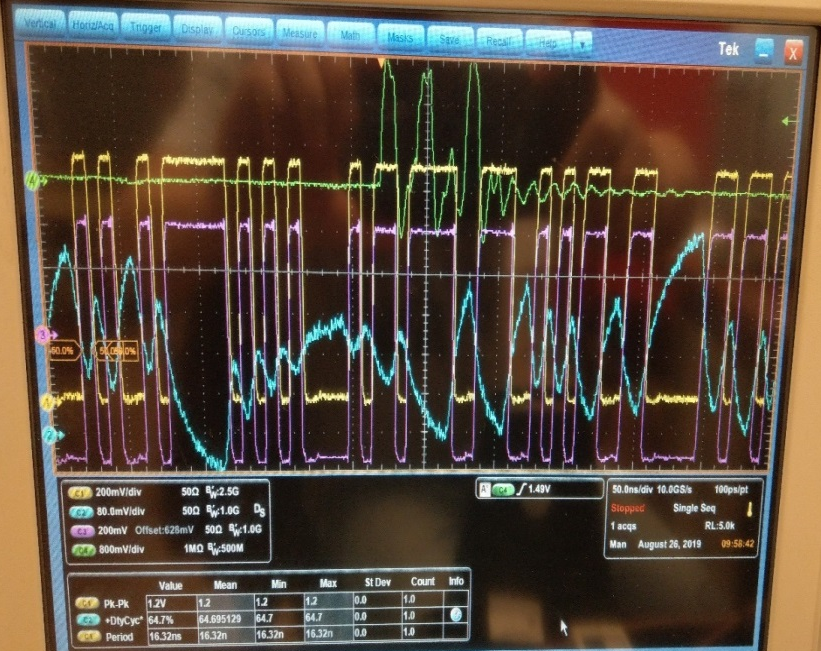}
\caption{The green one is the error indicator from the prbs checker; yellow one is the received prbs data (after equalizer) after 100 meter Ethernet cable; pink one is a delayed version of the original prbs data, it has been delayed for 500 ns to be able to compare with the received data after the long cable; the blue one is the received prbs data before equalizer, on which we can clearly see the attenuation effect.}
\label{fig:eror_capture}
\end{figure}

In the picture \ref{fig:eror_capture}, we can see the 4 channels that are used. The 3 digital waveforms show 3 signals from the FPGA: the green one is the error indicator from the prbs checker; the yellow one is the received prbs data (after equalizer) after 100 meters Ethernet cable; the pink one is a delayed version of the original prbs data, it has been delayed for 500 ns to be able to compare with the received data after the long cable; the blue one is the received prbs data before equalizer, where we can clearly see the attenuation effect. In this setup, we injected an artificial error into the original prbs data stream. We can see from the green waveform that the error has been captured successfully. By comparing the delayed original data with the received one after equalizer, we can tell that they are fully matching. Based on this observation, we can get such inference: when an error has been captured, if the received data after equalizer matches the delayed original data, it means that the error is from the data source, otherwise we can conclude that the error is from the external noise.\\
\begin{figure}[t!]
\centering
\includegraphics[width=2.3in]{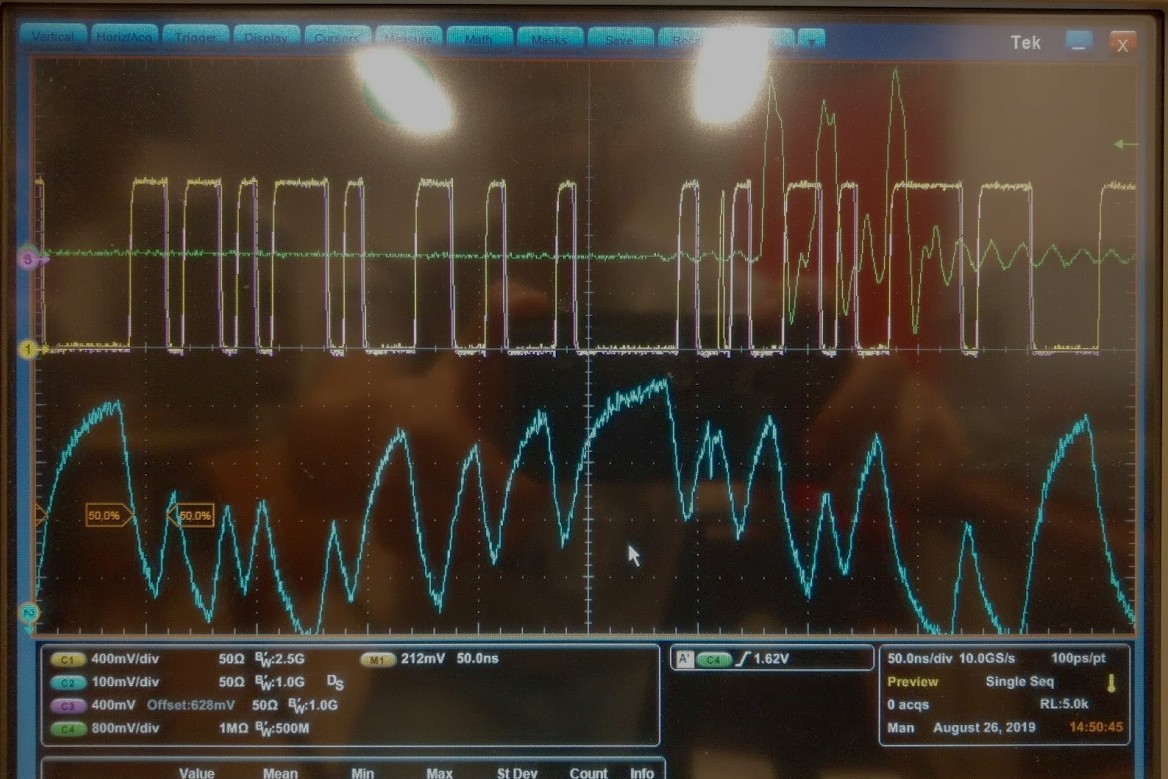}
\caption{The errors are captured during a long term test. In order to have a clear view, we have moved the delayed version of the original data to overlap with the received data after equalizer.}
\label{fig:eror_capture2}
\end{figure}
The channel settings in Figure \ref{fig:eror_capture2} are the sames as the one used in the previous picture. The errors are captured during a long term test. In order to have a clear view, we have moved the delayed version of the original data to overlap with the received data after equalizer. It is very easy to see that there is a difference between them: a glitch which lead to the error was not seen in the original data, while it can be seen in the received data before the equalizer. We can thus conclude that the error is due to the external noise.

\begin{figure}[t!]
\centering
\includegraphics[width=2.3in]{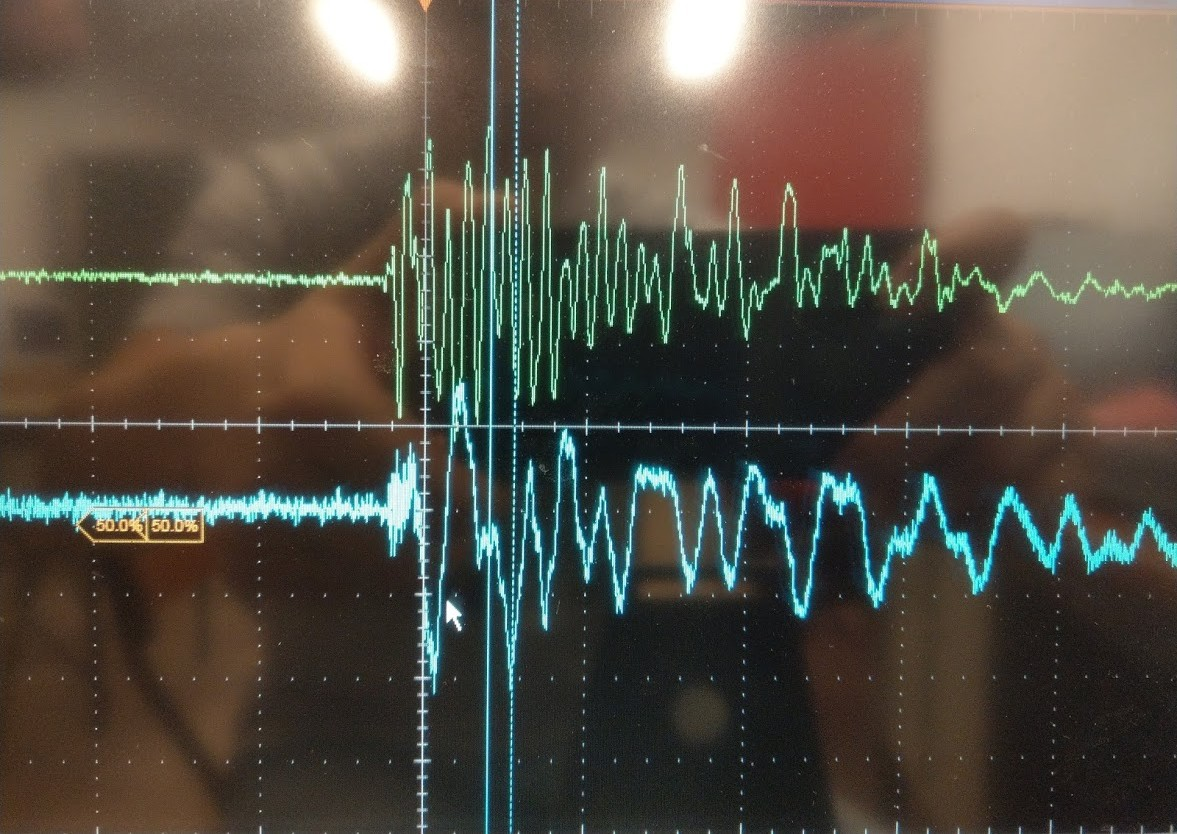}
\caption{The green one is coming from a SMA cable connecting to the TTIM, represents the error indicator of the prbs checker, the blue one is a differential probe connecting to a 100 ohm terminator, by plugging a load in the adjacent power socket as the TTIM power supply, an Electrical Fast-Transient and its impact has been emulated.}
\label{fig:eror_capture3}
\end{figure}
In order to further study the external noise impact, an experiment was performed, which is shown in the Figure \ref{fig:eror_capture3}. The green curve is coming from a SMA cable connecting to the TTIM, and represents the error indicator of the prbs checker. The blue one is a differential probe connecting to a 100 Ohm terminator. By plugging a load in the adjacent power socket as the TTIM power supply, the impact of Electrical Fast-Transient has been emulated. We can conclude that Electrical Fast-Transient are one of the noise sources which can cause data transfer error. The final performance of the system will be highly related to the noise level of the real experiment environment.

Since a perfect shielding is not possible, we can most likely hardly avoid bit error during data transfer if we continue to use Ethernet cables as the transfer media. A valid mitigation method should be implemented as an error correction code into the communication protocol.

\section{Conclusion}
We have presented an automatic test system of the BEC system of the JUNO experiment. The BEC system is used to link the PMT signals from the underwater boxes to the trigger system and to transmit the system clock and triggered signals. Each BEC connects to 48 underwater boxes, and in total around 150 BECs are needed. We have built an automatic test system to check the physical link performance of the BEC, before applying the real connection with underwater system.

The test system is based on a custom designed FPGA board, with a JTAG interface to the PC. The system can generate and check different data patterns at different speeds for 96 channels simultaneously. The test results of 1024 continuously clock cycles are automatically uploaded to the PC periodically. The setup of the automatic test system of the BEC is described and the latest test results are presented.

The current setup is easy to operate and can be used for mass production tests. A future work is to redo the tests with a better time resolution. To achieve this, we plan to use IPbus based firmware \cite{Larrea_2015} and Python scripts. Finally, we will redo the test on a setup that will include the complete trigger chain.

\bibliographystyle{unsrt}
\bibliography{tns}
\end{document}